# Real-time COVID-19 hospital admissions forecasting with leading indicators and ensemble methods in England

Jonathon Mellor[1]*, Rachel Christie[1], Robert S Paton[1], Rhianna Leslie[1], Maria Tang[1], Martyn Fyles[1], Sarah Deeny[1], Thomas Ward[1], Christopher E Overton[1,2]

1. UK Health Security Agency, Data Analytics and Science, Noble House, London, United Kingdom
2. University of Liverpool, Department of Mathematical Sciences, Liverpool, United Kingdom

*Corresponding Author: Jonathon.Mellor@UKHSA.gov.uk

## Abstract

### Background

Hospitalisations from COVID-19 with Omicron sub-lineages have put a sustained pressure on the English healthcare system. Understanding the expected healthcare demand enables more effective and timely planning from public health.

### Methods

We collect syndromic surveillance sources, which include online search data, NHS 111 telephonic and online triages. Incorporating this data we explore generalised additive models, generalised linear mixed-models, penalised generalised linear models and model ensemble methods to forecast over a two-week forecast horizon at an NHS Trust level. Furthermore, we showcase how model combinations improve forecast scoring through a mean ensemble, weighted ensemble, and ensemble by regression.

### Results

Validated over multiple Omicron waves, at different spatial scales, we show that leading indicators can improve performance of forecasting models, particularly at epidemic changepoints. Using a variety of scoring rules, we show that ensemble approaches outperformed all individual models, providing higher performance at a 21-day window than the corresponding individual models at 14-days.

### Interpretation

We introduce a modelling structure used by public health officials in England in 2022 to inform NHS healthcare strategy and policy decision making. This paper explores the significance of ensemble methods to improve forecasting performance and how novel syndromic surveillance can be practically applied in epidemic forecasting.

## 1. Introduction

Over the course of 2022 there were over 390,000 hospitalisations due to COVID-19 in England, an increase of approximately 90,000 from the year before [1]. This was a consequence of a reduction in non-pharmaceutical interventions and the high infectivity of the Omicron sub-lineages compared to previous variants [2]. The burden on healthcare systems remains high and hospital admissions with COVID-19 are a key metric for monitoring the SARS-CoV-2 pandemic. While the infection hospitalisation risk has reduced since 2021 [3] the higher transmission of Omicron and its emergent sub-lineages has sustained epidemic waves of admissions from COVID-19 in England and worldwide. These admissions are primarily in older age groups [4], and those with comorbidities [5]. Once admitted, patients with COVID-19 occupy beds for a median of 7.0 days in 2022 [6] with variation due to regional heterogeneity, risk factors and the patient pathways taken [7].

Due to the healthcare burden of COVID-19, system leaders request hospital admissions forecasts to inform management and policy decisions. There are a range of existing COVID forecasting approaches and models [8], as for epidemiological forecasting more generally [9], though they have limitations for our specific policy problem. Mechanistic or transmission models rely on parametric values, such as relative susceptibility in a population [10] which are often unknowable for new variant-driven waves and can change substantially over time. On the other hand, purely time series models, such as ARIMAs, will not be able to anticipate turning points such as epidemic peaks [11], which is the period where accurate forecasts are crucial. To enhance performance, leading indicators such as incidence can be incorporated to help predict changes in hospital metrics [12], though each data stream is subject to its own biases and sources of error and may have a changing relationship with hospitalisations over time [13]. Due to Universal testing in the community ending in 2022 [14] there is a greater reliance on non-clinical leading indicators and novel syndromic surveillance in order to anticipate hospital admissions. There has been significant work on the analysis of leading indicators of COVID-19 activity [15, 16], but limited exploration across Omicron epidemic waves. There is significant body of work that shows forecasting accuracy can be improved by bringing together a range of model structures in an ensemble [17], for example using an unweighted average of candidate forecasts [18].

In this paper we introduce multiple model structures used to forecast hospital admissions in England throughout 2022 into 2023 operationally in UKHSA – which we validated across multiple epidemic waves. These models rely on a single time series or utilise leading indicators to forecasts admissions at National Health Service (NHS) Trust level - a collection of hospitals. These projections are produced at NHS Trust, NHS Commissioning Region and national levels in England. We both combine data for individual models and combine models in ensembles [19], using three different methods. This reduces the bias of individual models to improve predictive performance. Importantly, we show how these models score over time and contrast the different approaches and their performance throughout the epidemic wave, using proper scoring rules [20].

## 2. Methodology

### 2.1 Data

### 2.1.1 Hospital Admissions

NHS England (NHSE) COVID-19 data is provided by individual acute NHS Trusts in England, who deliver a daily situation report (SitRep) covering the previous 24 hours on metrics relating to patients, beds, and staff [21]. The data records the number of new patients and inpatients in the past 24 hours with a laboratory-confirmed positive COVID-19 test [22]. We define a COVID-19 admission as any patient who tested positive before admission or within their first 2 days of arrival - we are interested in community acquired admissions, so our definition excludes expected hospital acquired infections.

### 2.1.3 Geographic Structure

The NHS in England is structured hierarchically, with national oversight from NHS England and seven commissioning regions. The hospitals within each commissioning region are managed as organisational units called NHS Trusts, each Trust with secondary care responsibility may have one or many acute / emergency hospitals. The NHS Trusts cross administrative boundaries, with nearby Trusts serving overlapping populations. This hierarchical structure can be incorporated into modelling and is shown visually in *Supplementary Figure A*.

### 2.1.3 Leading Indicators

Healthcare seeking behaviour may not lead hospitalisation at an individual linkable level, but we expect population level behaviour to lead aggregate admissions. For example, increases in Google Searches for “what are COVID symptoms” correlate with increased transmission in an area, which should cause increased hospitalisations in the nearby Trusts following some time delay. A probabilistic population mapping was created linking patient discharge locations in a lower tier local authority (LTLA) to a service provider (NHS trust), in a similar manner to the *covid19.nhs.data* R package [23]. We can then map trends in local populations healthcare seeking behaviour (recorded in administrative boundaries) to nearby NHS Trusts, as well as their population catchment sizes.

Candidate leading indicators were evaluated for both strength of statistical relationship with admissions, and the likelihood of being operationalisable [13]. Ultimately, the Google Trends syndromic search terms, and NHS 111 Pathways telephonic triage (calls and online), were selected due to strong correlations with localised clinical risk – originally explored in [24]. For Google, individual search terms monitored were combined by topic to increase robustness of signal. The NHS 111 Pathways were separated into online and calls data sources and aggregated to type of triage and age group.

## 2.2 Models

As there are multiple models discussed and combined in this manuscript, the high-level implementation of models used are outlined in Table 1.

| Model name | Model type | Data sources / model input (*) | Ensemble approach |
|---|---|---|---|
| Univariate Baseline | Generalised additive model | Hospital admissions | None |
| Univariate HGAM | Hierarchical generalised additive model | Hospital admissions | None |
| Google Trends | Penalised generalised linear model, input into generalised linear mixed model | Hospital admissions<br>Google syndromic surveillance | None |
| 111 Calls | Penalised generalised linear model, input into generalised linear mixed model | Hospital admissions<br>Syndromic telephonic triage | None |
| 111 Online | Penalised generalised linear model, input into generalised linear mixed model | Hospital admissions<br>Online syndromic telephonic triage | None |
| Combined Indicator | Penalised generalised linear model, input into generalised linear mixed model | Hospital admissions<br>Google syndromic surveillance<br>Syndromic telephonic triage<br>Online syndromic telephonic triage | Include data sources in the same model |
| Ensemble by mean | Ensemble | *Univariate HGAM<br>*Google Trends<br>*111 Calls<br>*111 Online | Mean of central estimate and quantiles |
| Ensemble by score | Ensemble | *Univariate HGAM<br>*Google Trends<br>*111 Calls<br>*111 Online | Weight models in average using previous prediction interval scores |
| Ensemble by regression | Ensemble | *Univariate HGAM<br>*Google Trends<br>*111 Calls<br>*111 Online | Determine weights by regression on central fits of previous predictions |

*Table 1. Breakdown of the different models discussed in this manuscript, their data sources, and how they relate to each other.*

### 2.2.1 Univariate

We use two univariate (hospital admissions time series as the only predictor) models in this study. The first, “Univariate HGAM”, is a Hierarchical Generalized Additive Model, which estimates and extrapolates the local growth rate per hospital Trust, with splines through time at both Trust and NHS Region levels. The second, “Univariate baseline”, has a similar structure, but is not spatially hierarchical, instead fitting splines through time for each Trust independently. As a simple to apply statistical model, we use the baseline GAM model throughout to compare with other methods. The models are fit regionally for computational efficiency, and the GAMs fit using the *mgcv* R package [25].

To forecast admissions, we need to model how the daily admission counts are changing over time, $H(t)$. On short timescales, epidemics can often be described using an exponential structure, where the incidence at time $t$ is a function of some initial incidence and exponential growth/decay for $t$ days. Assuming hospital admissions are linearly related to incidence, we have

$$H(t) = H(0)e^{rt},$$

where $r$ is the exponential growth rate. Over an epidemic, the growth rate is rarely constant. This model can be generalised using a smooth function of time $s(t)$ rather than $rt$ in the exponent, i.e.

$$H(t) = H(0)e^{s(t)}.$$

By fitting such a model to time-series data on hospital admissions, one can generate short-term forecasts by assuming that for all $t > t_{\max}$ the exponential growth rate remains constant, i.e., $s(t) = s(t_{\max}) + (t_{\max} - t)s_1$. Here $s_1$ is the instantaneous exponential growth rate at $t = t_{\max}$. Assuming the smooth function $s(t)$ is known, $s_1$ is approximately the first derivative of $s(t)$, evaluated at $t_{\max}$. This can be shown by taking a Taylor expansion of the smooth function,

$$s(t_{\max} + h) = s(t_{\max}) + h\left.\frac{ds}{dt}\right|_{t_{\max}} + \cdots \approx s(t_{\max}) + hs_1.$$

Substituting this back into our hospital admissions formula gives

$$H(t_{\max} + h) = H(0)e^{s(t_{\max})+hs_1} = H(t_{max})e^{hs_1}.$$

Hospital admissions data are noisy integer-valued counts, with stochasticity from both the epidemic spread and the likelihood of requiring medical care after infection. To model this integer-valued noise, we assume that observed hospital admissions are samples from a negative binomial distribution, with expected value $H(t)$. To fit this model, we use a Generalised Additive Model with logarithmic link function and negative binomial error structure. Under this, we obtain

$$log\left(H_{trust_i}(t)\right) \sim \beta_0 + R_{trust_i} + s_{trust_i}(t) + R_{wday}(t), (1)$$

where $\beta_0$ is an intercept, $R_{trust_i}$ a random effect on Trust $i$, and $s_{trust_i}(t)$ is a penalised cubic regression spline and $R_{wday(t)}$ is a random effect on the day-of-week at $t$. Using the penalised spline, the out of sample prediction for future dates assumes a linear relationship with time, with gradient equal to the first derivate of the spline at $t_{\max}$. Therefore, we can use out of sample prediction from the GAM to forecast admissions using a continued exponential trend.

Baseline GAM

The baseline GAM model is obtained by fitting Equation (1) to data from individual NHS Trusts independently. This leads to a unique spline for each Trust.

Univariate hierarchical GAM

The baseline GAM leads to very high uncertainty at Trust level and assumes each Trust $i$ has an independent trend, which is typically not the case for epidemics, where spatial correlation is usually strong. Therefore, we instead construct a hierarchical GAM that accounts for correlation between Trusts nested within NHS Regions. We consider the structure

$$log\left(H_{trust_i}(t)\right) \sim \beta_0 + R_{\mathrm{trust_i}} + s_{\mathrm{trust_i}}(t) + s_{\mathrm{region_i}}(t) + R_{\mathrm{wday}}(t).$$

We run the model for each region independently. For the Trust splines, we use a hierarchical structure based on [26]. Since the regional models are independent, this nests the Trusts within regions. The regional spline captures the average trend across the region, with the Trust level splines and random effect $R_{\mathrm{trust_i}}$ adding trust level variation.

### 2.2.2 Leading indicator models

Each leading indicator model "Google Trends", NHS "111 Calls" and "111 Online" use a penalised generalised linear model (pen-GLM) to fit a smoothed admissions response variable with the leading indicators as predictors - then a generalised linear mixed effect model (GLMM) to fit directly to the data using the pen-GLM output as a predictor. We do this to capture the trends within the highly stochastic indicators and admissions data at fine spatial scales, which performed better than modelling the data directly within one model in initial exploration.

The leading indicators, denoted by $x_t$, are noisy at fine spatial scales, as are hospital admissions, therefore the pen-GLM uses smoothed (via LOESS, given by $u(x_t)$) indicators to predict smoothed admissions. The relationship between leading indicator time series is estimated at national and regional levels, to allow for spatial variation in leading relationships and national trends.

To construct the regression, a fixed lag was introduced between the indicator and admissions by the forecast horizon $h$ steps. This allowed a prediction of admissions at $H(t = t_{max} + h)$ using leading indicators at $x_{t=t_{max}}$. As the optimal time-delay aligning indicator and admissions series is unknown a priori of an epidemic wave, we add further lags $l$ between the two series, at $t = -h - l$, with the maximum plausible lag at $l_{max}$. This inclusion of further lags allows a higher chance of capturing a correlation in the model, though this comes at the cost of a highly autocorrelated regression. Across the $J$ indicators indexed by $j$ and the catchment population size of the Trust, $p_i$, the model across the country becomes

$$\begin{aligned} &log\big(u(H_{\mathrm{trust}_i}(t))\big) \\ &\qquad = \beta_{\mathrm{region}_i} + \beta_{\mathrm{trust}_i} + \log{(p_i)} + \sum_{0}^{l_{max}} \sum_{0}^{J} \beta_j \, u(x_{t-h-l,j}) \\ &\qquad + \sum_{0}^{l_{max}} \sum_{0}^{J} \beta_{j,\mathrm{region}_i,l} u(x_{t-h-l,j}) \ \mathrm{region}_i. \end{aligned}$$

This produces many $\beta$, which we can penalise out to reduce collinearity, improve performance and exclude poor performing indicators and lags. We perform penalisation using a LASSO method, implemented using the R package *glmnet* [27] and a negative binomial error structure.

However, this does not predict admission counts directly - only the smoothed trend, which introduces bias and does not allow generation of prediction intervals. Therefore, we use the output of this model, denoted by $\widehat{H}(t)$ as a covariate in a GLMM with admission counts as the response variable, using the structure

$$log\left(H_{trust_i}(t)\right) =$$

$$\beta_0 + \beta_{trust_i}$$

$$+ \beta_1 \, 1_{\{t<t_{max}-c\}}(t) \; log\left(\widehat{H}(t)\right)$$

$$+\beta_2 \, 1_{\{t\geq t_{max}-c\}}(t) \; log\left(\widehat{H}(t)\right)$$

$$+\text{wday}(t).$$

Where $1_{\{X\}}$ is an indicator function. This allows different coefficients of $log\left(\widehat{H}(t)\right)$ to be estimated, allowing for a correction near $t_{max}$, where $c$ is an integer value of days, taken as $c = 14$. The package *mgcv* is used for this GLMM, as with the GAMs a negative binomial error structure is assumed and modelled at an NHS Commissioning Region level.

### 2.2.3 Prediction intervals

To calculate prediction intervals from the fitted GAM and GLMM models, we need to capture both parameter uncertainty and the uncertainty in the error structure of the data generating process. From the fitted models, we generate a posterior distribution of the model parameters by assuming the coefficients are distributed according to a multivariate normal with mean equal to the central estimates of the model and variance-covariance matrix defined by the fitted model. From the fitted model, we capture uncertainty in parameter estimates by simulating from a multivariate normal, sampled 2000 times. For each parameter sample we produce a model forecast, and then produce estimates of the uncertainty in the mean of the forecast. These forecasts, however, do not capture the noise in the data generating process. To capture this, we take each forecast sample and simulate a sample from a negative binomial distribution with the forecast sample as the mean and theta parameter taken from the fitted GAM/GLMM. Therefore, for each posterior sample of the coefficients, we have a corresponding sample from the data generating process. Aggregating these sample trajectories from the data generating process, we can calculate prediction intervals.

Since the GAM and GLMM model use a hierarchical structure nesting Trusts within Regions, we can produce calibrated prediction intervals at both Trust and Region level. At Trust level, we generate forecast samples for each Trust and then simulate the negative binomial noise. At Region level, we aggregate the forecast samples from each nested Trust to Region level. Taking the Region level forecast samples, we then simulate the negative binomial noise using this as the expected value. Since each Region is run independently, we do not have calibrated prediction intervals at Nation level. For operational purposes, we can aggregate the prediction intervals across each Region, but we do not score these results since these will be uncalibrated.

### 2.2.4 Ensembles

There is substantial literature on how forecasting, and specifically epidemic forecasting, can be improved by ensembling multiple models together. In this manuscript we compare three such methods. We use a common ensemble method “ensemble by mean” comparing it to two methods which leverage past predictive performance with the aim of improving accuracy above this simple average. Ensembles have been shown to improve predictive performance for large national multi-team [28] [29], though in contrast the models we use are all data driven and produced from a single team. The models selected for inclusion were chosen to each tackle specific short comings of the other candidate models. The Univariate HGAM, as a growth rate extrapolation model, performs well in epidemic growth and decline phases, but the extrapolation fails at first order turning points. The leading indicator models do not capture the epidemic dynamics as well but can anticipate short term changes in turning points. However, as the indicators are not consistent, multiple data sources are used to minimise the risk of spurious prediction and increase operational resilience.

The first method “ensemble by mean”, for Trust $i$ at time $t$, is given by

$$\hat{y}_{i,t} = \frac{1}{M}\sum_{m=0}^{M} \hat{y}_{i,t,m}$$

where each individual model is noted by $m$ and there are $M$ models in the ensemble.

The second ensemble approach utilises the scores of forecasts run on historic data to weight predictions, in this case the forecasts are run weekly – “ensemble by score”. For a forecast into the future, we index by week $b = 0$, for a given week we can therefore use how the model performed in the preceding $b - 1$ week. We can, of course, only determine a weighting using information that would be available at $b = 0$, so for a forecast horizon of $h = 14$ at $b = -1$, we must truncate to $t \leq 7$ for the weekly case. To create this ensemble on week $b$, we evaluate the model performance at $b - 1$, and produce a weighting for the relative score of each candidate model in the ensemble. We use the Weighted Interval Score, calculated through the *scoringUtils* R package [30] to measure historic performance, we use this package to evaluate all forecasts throughout this manuscript.

The average weighted interval for a model is then given by interval score function $wis(y)$

$$q_{m,b} = \sum_{t=0}^{7}\sum_{i=0}^{I} wis(\hat{y}_{i,t,m,b-1})$$

We get the weighting of each model as

$$w_{m,b} = \frac{q_{m,b}}{\sum_{m=0}^{M} q_{m,b}}$$

And therefore, a prediction of

$$\hat{y}_{i,t,b} = \sum_{m=0}^{M} \hat{y}_{i,t,m,b} \ q_{m,b}$$

The final ensemble method, "ensemble by regression", seeks to find the optimal combination of models through, in the simple case, an ordinary least squares. This structure is similar to the ensemble by score, as it uses the historic model performance on weeks $b - 1$, however, instead of directly scoring each model, the regression model finds a linear combination of the individual model's predictions that best estimate the historic data. Using the idea that past performance/model weighting can inform future best weighting, we define a regression of

$$y_{i,t,b-1} = \sum_{m=0}^{M} \beta_{\mathrm{m,b-1}} \sum_{t=0}^{7} \sum_{i=0}^{I} \hat{y}_{i,t,m,b-1}$$

From this we can create a weighted ensemble for week $b$ extracting the regression coefficients $\beta_{m,b-1}$ for each model

$$\hat{y}_{i,t,b} = \sum_{m=0}^{M} \hat{y}_{i,t,m} \times \beta_{m,b-1}$$

We extend this further to the Bayesian case, where we have prior estimates on the values of $\beta_{m,b-1}$. We take a normal prior on the weighting of the models with a value $\beta_{m,b-1} = \frac{1}{M}$, 0.25 in this ensemble. This Bayesian framework allows us to have a prior belief in the best ensemble weighting, reconciling it with the best combination according to the data. For our model ensemble, a prior scale of 0.01 is shown, with the sensitivity shown in overall for BA.4/5 Supplementary Figure B and over time in Supplementary Figure C. Where models rely on previous week's predictions to score, the first week of data in the time series cannot be used. For this reason, the first week of data is excluded from all scoring.

## 3. Results

### 3.1 Epidemic Curves

In this study we showcase the modelling approach focusing on the 2022 Omicron BA.4/5 wave, with further investigation for the 2022/23 Winter wave provided in the supporting documentation. The start date/end date and how they were defined for the BA.4/5 and Winter 2022 waves is given in *Table 2*. The national shapes of the epidemic waves are shown in *Figure 1.* Regional breakdowns of the epidemic curves are shown in *Supplementary Figure D*.

| **Wave** | **Phase** | **Week Start** | **Week End** |
|---|---|---|---|
| BA.4/5 | Growth | 2022-05-15 | 2022-06-05 |
| | Peak | 2022-06-05 | 2022-06-19 |
| | Decline | 2022-06-19 | 2022-09-11 |
| Winter 2022 | Growth | 2022-11-13 | 2022-12-18 |
| | Peak | 2022-12-18 | 2022-01-01 |
| | Decline | 2022-01-01 | 2023-01-22 |

*Table 2. The defined start and end dates of the waves and epidemic phases explored for forecasts. The methodology for defining the wave start and end dates is given in Supplementary Section A.*

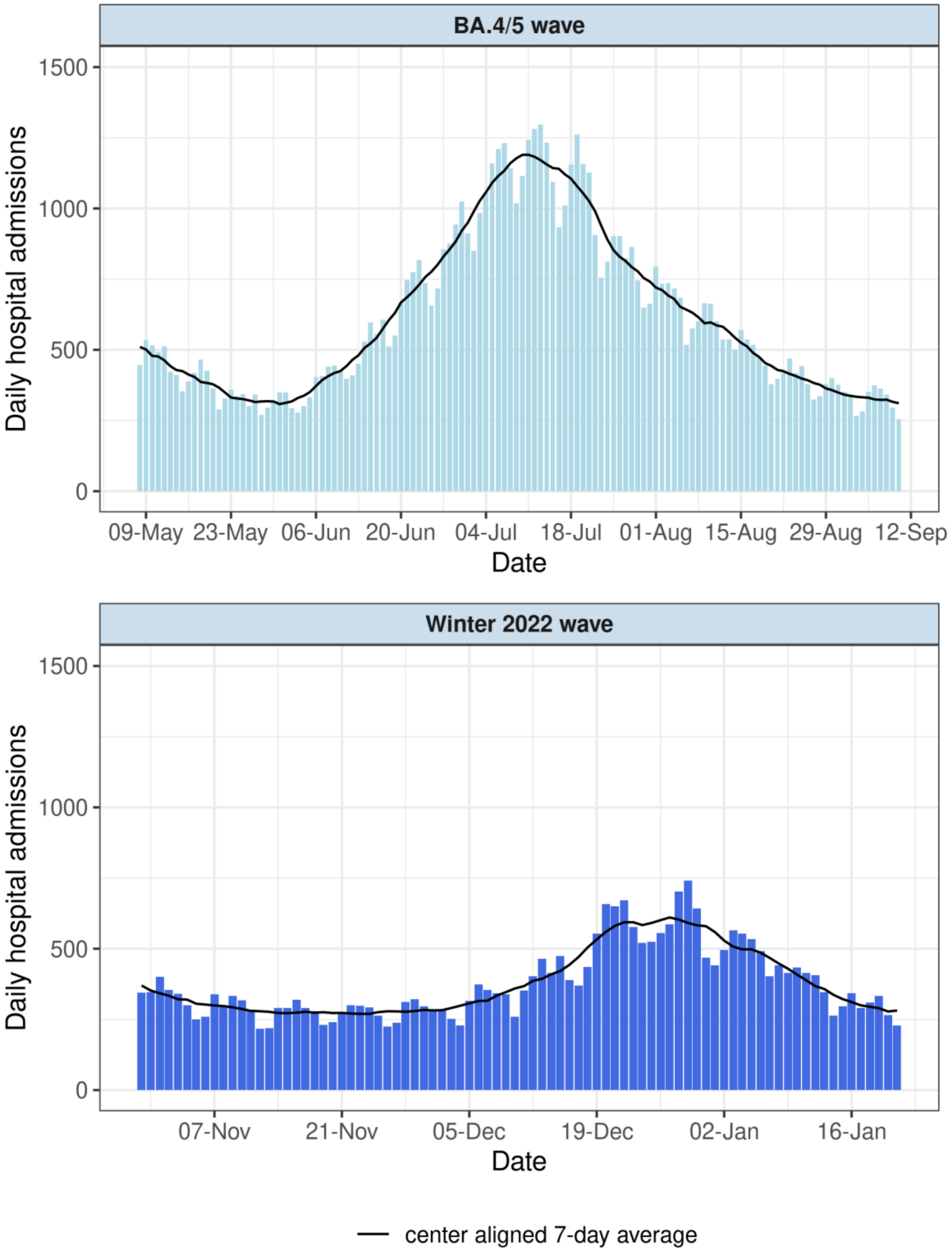

*Figure 1. The hospitalisation epidemic curves for the Omicron BA.4/5 and Winter 2022 waves. The BA.4/5 wave peaked at over 1,200 admissions per day and lasted through to trough approximately 14 weeks, compared to 600 and 10 weeks respectively in the Winter wave. The BA.4/5 rises fast following its low turning point compared to its slow decay. The Winter wave's smaller peak rises slowly from the baseline daily admissions around 250. Both waves show strong day-of-week effects in reported counts.*

### 3.2 Forecast performance over time

Example forecasts for the BA.4/5 wave are shown in Figure 2. At a national level, we can see that the models based on hospital admissions alone (Univariate baseline, Univariate HGAM) over predict at the national peak. This effect is much smaller for the leading indicator-based models, which mostly avoid overprediction at the peak – though they do struggle to increase fast enough in the growth phase of the wave. All models appear to predict the decline phase well. In the ensemble models we see a mix of the univariate behaviour of over-predicting at peaks, though this effect is muted to different degrees. Similar effects are shown in *Supplementary Figure E* for the Winter 2022/23 wave, though the leading indicator models do not rise as fast in the weeks preceding the peak.

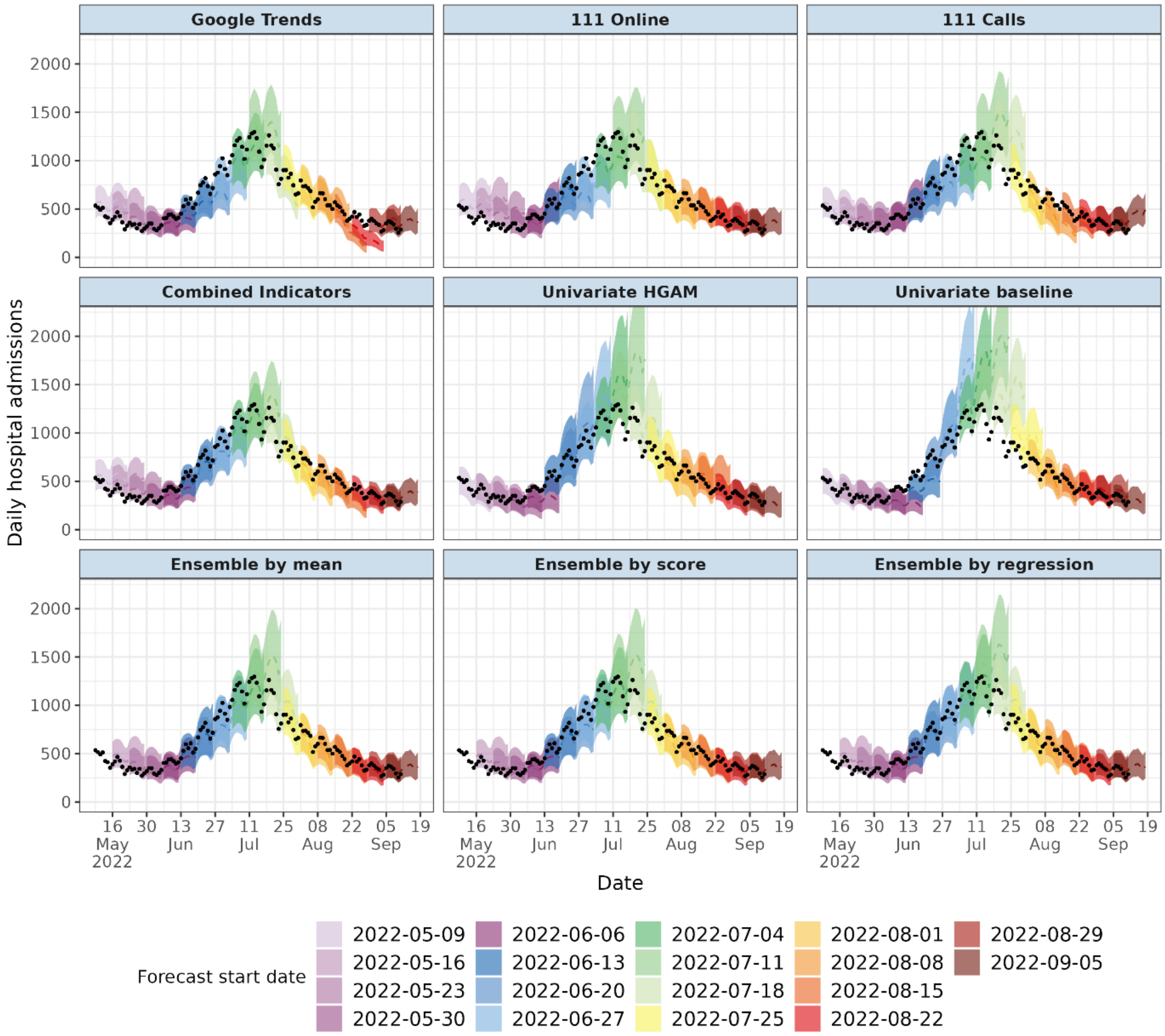

*Figure 2. The example forecasts of the different model structures for the BA.4/5 wave for each week period. The regional forecasts from the GAMs are aggregated to national level to show the epidemic curve and represent forecasts. The corresponding figure for the Winter 2022/23 wave is given in Supplementary Figure E.*

It’s important to analyse forecasts not just overall, but at specific time points to understand when they perform well or poorly, particularly when the underlying trend is an epidemic. The metrics over time for individual (non-ensembled) models are shown in *Figure 3*. The first metric, the interval score, gives a measure of model error, sharpness, and calibration - for this metric lower values indicate better performance. The second, bias, indicates whether the model is over or under predicting on average – where the closer to zero the better performance. The final metric, 95% coverage, tells us how many of the true values are contained within the prediction intervals of the models, so the nearest to 95% is best. The weekly national hospitalisation ratio $\frac{H(t)}{H(t-7)}$ is shown to indicate growth / decline phases. In *Figure 3*, we can see across all models the interval score is highest for the forecast including the peak, and broadly follows the pattern of the hospitalisation ratio across all models. All individual models perform similarly in the decline phase, which from the ratio has a constant decline rate. The univariate models have the highest interval scores at the epidemic peak, and we can see from their bias that they are substantially overshooting the turning point – as expected due to their model structure. The bias in the growth phase differs between the indicator and univariate models, with the indicators underpredicting and the univariates overpredicting, indicating performance could be improved by ensembling.

We extend these individual models using a variety of ensemble methods, and as shown in *Figure 4*, all the ensembles’ interval scores (top metric) outperform the individual models across all time periods. Across time the ensemble by mean and ensemble by score perform similarly in all metrics with near identical interval scores, as expected. This is because the weighting method with scoring approaches equal weighing when models perform similarly. This is shown in *Figure 4* as the top metric – the black and green lines are at the same position. The bias (middle metric) for the different ensemble approaches falls between the Univariate HGAM and Combined Indicator, and is generally closer to zero, showing the ensembles are effectively reducing bias from the individual models. The ensemble by regression has higher interval score (poorer performance) than other ensemble approaches at the peak, due to higher weighting of the univariate model, shown in further detail in *Supplementary Figure F*. The 95% coverage of the ensemble models drop slightly in the growth phase; however, this is small in comparison to the individual models.

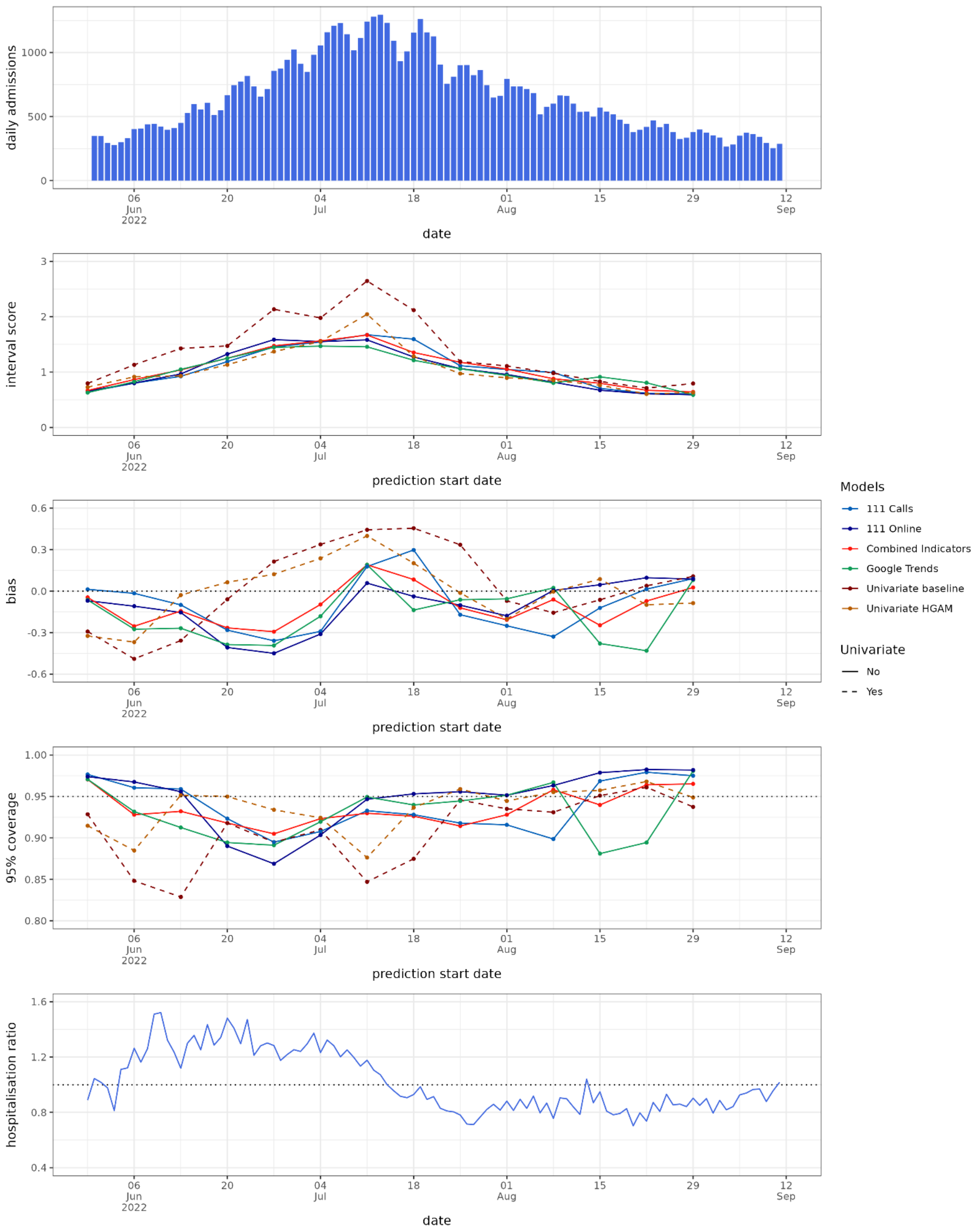

*Figure 3. Performance of individual models (non-ensembled) over time for the BA.4/5 wave. The epidemic curve (top) and hospitalisation ratio (bottom), the admissions divided by the admissions seven days prior, are shown to contextualise scores. The prediction start date represents the first date of prediction, where the predictions will be on the subsequent h=14 days. Supplementary Figure G contains the equivalent metrics for the Winter 2022/23 wave.*

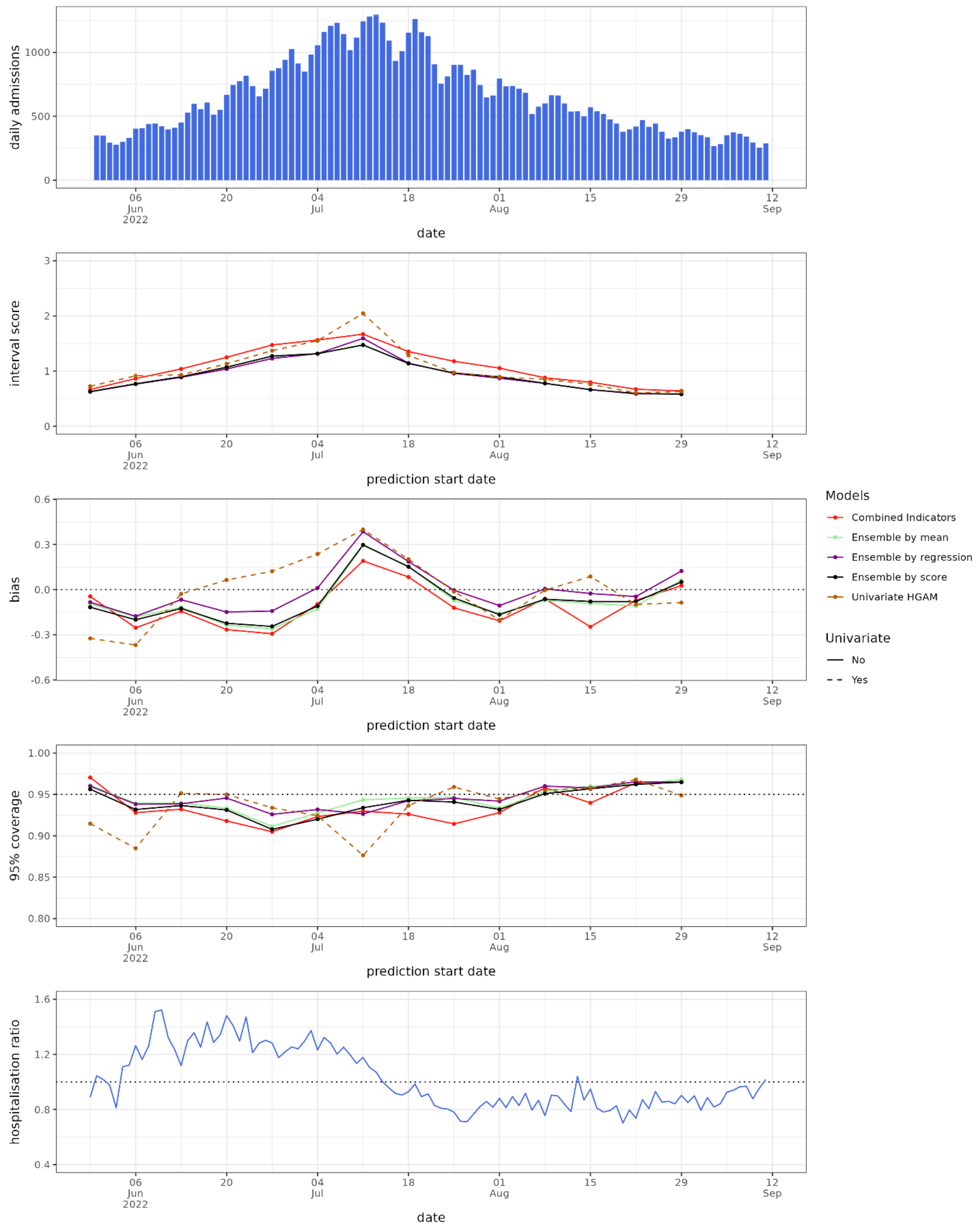

*Figure 4. Performance of ensemble models over time for the BA.4/5 wave, the Univariate HGAM and Combined Indicators model are included to compare performance. The epidemic curve (top) and hospitalisation ratio (bottom), the admissions divided by the admissions seven days prior, are shown to contextualise scores. The prediction start date represents the first date of prediction, where the predictions will be on the subsequent h=14 days. Supplementary Figure H contains the equivalent metrics for the Winter 2022/23 wave.*

### 3.3 Overall performance and forecast horizons

In *Table 3* we show how the models scored overall in a wave. Whilst instantaneous performance within wave is important, we need to understand overall how models performed, stratified by length of forecast horizon, $h$. This is shown at NHS Trust level forecast for the BA.4/5 wave, though the Region waves are shown in *Supplementary Tables A and B*. In *Table 3*, for the interval score and median average error the ensemble models outperform all other models, though there is no clear best approach in this case. Unexpectedly, the Combined indicator model performs worse than the individual indicators. This is perhaps due to the larger feature space of indicator variables with high collinearity – which the penalisation parameter may not be tuned strongly enough to adjust for. The model may struggle to select an optimal combination of features, putting too much weight on non-informative variables. The ensemble models perform well on interval score and error in comparison to the individual models of larger forecast horizons. We can see from the table that on average the univariate models overpredict, the leading indicators underpredict, particularly in the growth phase of a wave. For the Winter 2022/23 wave the interval scores and median absolute error are lower than for the BA.4/5 wave, shown in *Table 4*, though the waves are different epidemic shapes - the Winter 2022/23 wave may be easier to predict due to its flatness and smaller peak – shown in *Figure 1*.

| **Trust geography** | | | | | | |
|---|---|---|---|---|---|---|
| **model** | **forecast horizon** | **interval score** | **95% coverage** | **median absolute error** | **underprediction** | **overprediction** |
| Combined Indicators | 7 | 1.02 | 0.945 | 2.31 | 0.533 | 0.323 |
| Univariate baseline | 7 | 1.08 | 0.940 | 2.45 | 0.441 | 0.461 |
| Ensemble by mean | 7 | 0.875 | 0.953 | 2.00 | 0.436 | **0.280*** |
| Ensemble by regression | 7 | **0.870*** | 0.953 | **1.99*** | **0.391*** | 0.314 |
| Ensemble by score | 7 | 0.873 | **0.950*** | 2.00 | 0.433 | 0.281 |
| Google Trends | 7 | 0.955 | 0.948 | 2.17 | 0.522 | **0.280*** |
| 111 Calls | 7 | 0.979 | 0.951 | 2.22 | 0.506 | 0.312 |
| 111 Online | 7 | 0.974 | 0.955 | 2.21 | 0.503 | 0.302 |
| Univariate HGAM | 7 | 0.897 | 0.955 | 2.04 | 0.399 | 0.344 |
| Combined Indicators | 14 | 1.070 | 0.938 | 2.42 | 0.556 | 0.350 |
| Univariate baseline | 14 | 1.330 | 0.913 | 2.95 | 0.492 | 0.635 |
| Ensemble by mean | 14 | 0.919 | 0.944 | **2.10*** | 0.456 | 0.302 |
| Ensemble by regression | 14 | 0.918 | 0.945 | **2.10*** | **0.405*** | 0.346 |
| Ensemble by score | 14 | **0.916*** | 0.941 | **2.10*** | 0.454 | 0.301 |
| Google Trends | 14 | 1.020 | 0.933 | 2.28 | 0.585 | **0.279*** |
| 111 Calls | 14 | 1.040 | 0.940 | 2.35 | 0.540 | 0.343 |
| 111 Online | 14 | 1.030 | **0.947*** | 2.33 | 0.541 | 0.321 |
| Univariate HGAM | 14 | 1.020 | 0.939 | 2.30 | 0.420 | 0.435 |
| Combined Indicators | 21 | 1.17 | 0.920 | 2.60 | 0.621 | 0.387 |
| Univariate baseline | 21 | 1.75 | 0.885 | 3.77 | 0.564 | 0.936 |
| Ensemble by mean | 21 | **0.986*** | 0.934 | 2.25 | 0.477 | 0.342 |
| Ensemble by regression | 21 | 0.994 | **0.935*** | 2.27 | **0.418*** | 0.398 |
| Ensemble by score | 21 | 0.976 | 0.932 | **2.22*** | 0.474 | 0.334 |
| Google Trends | 21 | 1.11 | 0.916 | 2.44 | 0.690 | **0.271*** |
| 111 Calls | 21 | 1.10 | 0.934 | 2.46 | 0.560 | 0.375 |
| 111 Online | 21 | 1.16 | 0.930 | 2.59 | 0.592 | 0.392 |
| Univariate HGAM | 21 | 1.24 | 0.919 | 2.73 | 0.458 | 0.591 |

*Table 3. Scores of each individual and ensemble model across a range of forecast horizons averaged over the BA.4/5 waves, shown for predictions at NHS Trust level. The same scores are shown for regional predictions in Supplementary Table A. Best performing models within forecast horizon and metric are denoted with an asterisk (*).*

| **Winter 2022/23 - Trust geography** | | | | | | |
|---|---|---|---|---|---|---|
| **model** | **forecast horizon** | **interval score** | **95% coverage** | **median average error** | **underprediction** | **overprediction** |
| Combined Indicators | 7 | 0.746 | 0.956 | 1.69 | 0.384 | 0.242 |
| Univariate baseline | 7 | 0.801 | 0.945 | 1.81 | 0.353 | 0.32 |
| Ensemble by mean | 7 | 0.674 | 0.952 | **1.54*** | 0.334 | 0.224 |
| Ensemble by regression | 7 | 0.674 | 0.955 | **1.54*** | **0.301*** | 0.252 |
| Ensemble by score | 7 | **0.672*** | **0.951*** | **1.54*** | 0.334 | 0.223 |
| Google Trends | 7 | 0.680 | 0.959 | **1.54*** | 0.384 | **0.189*** |
| 111 Calls | 7 | 0.759 | 0.955 | 1.72 | 0.364 | 0.275 |
| 111 Online | 7 | 0.692 | 0.970 | 1.58 | 0.347 | 0.222 |
| Univariate HGAM | 7 | 0.697 | 0.953 | 1.58 | 0.327 | 0.257 |
| Combined Indicators | 14 | 0.831 | 0.938 | 1.85 | 0.443 | 0.269 |
| Univariate baseline | 14 | 1.000 | 0.923 | 2.23 | 0.391 | 0.465 |
| Ensemble by mean | 14 | **0.729*** | 0.939 | **1.66*** | 0.375 | 0.239 |
| Ensemble by regression | 14 | 0.735 | 0.943 | 1.68 | **0.336*** | 0.277 |
| Ensemble by score | 14 | **0.729*** | 0.937 | **1.66*** | 0.374 | 0.240 |
| Google Trends | 14 | 0.745 | 0.946 | **1.66*** | 0.429 | **0.207*** |
| 111 Calls | 14 | 0.801 | 0.940 | 1.78 | 0.436 | 0.250 |
| 111 Online | 14 | 0.744 | **0.951*** | **1.66*** | 0.408 | 0.221 |
| Univariate HGAM | 14 | 0.828 | 0.938 | 1.85 | 0.353 | 0.353 |
| Combined Indicators | 21 | 0.978 | 0.904 | 2.07 | 0.542 | 0.317 |
| Univariate baseline | 21 | 1.380 | 0.891 | 2.95 | 0.448 | 0.734 |
| Ensemble by mean | 21 | **0.795*** | 0.921 | 1.77 | 0.415 | 0.263 |
| Ensemble by regression | 21 | 0.813 | 0.924 | 1.82 | **0.376*** | 0.311 |
| Ensemble by score | 21 | 0.796 | 0.917 | 1.77 | 0.415 | 0.264 |
| Google Trends | 21 | 0.824 | **0.932*** | 1.79 | 0.458 | 0.254 |
| 111 Calls | 21 | 0.820 | 0.927 | 1.77 | 0.512 | **0.202*** |
| 111 Online | 21 | 0.808 | 0.929 | **1.73*** | 0.497 | 0.203 |
| Univariate HGAM | 21 | 1.060 | 0.911 | 2.30 | 0.395 | 0.523 |

*Table 4. Scores of each individual and ensemble model across a range of forecast horizons averaged over the Winter 2022/23 wave, shown for predictions at NHS Trust level. The same scores are shown for regional predictions in Supplementary Table B. Best performing models within forecast horizon and metric are denoted with an asterisk (*).*

## 4. Discussion

To improve forecasting capability at both local and national level in England, we developed a novel forecasting framework, based on generalised additive models. These models are flexible and fast, allowing them to be used in real-time and rapidly adjust to changing variant/immunological dynamics. In addition to just modelling time series trends, our framework allows the incorporation of syndromic surveillance data as a leading indicator, including Google Trends data and NHS 111 (non-emergency) data, which we show improves forecasting performance at epidemic changepoints.

The primary strength of our methods came from the ensemble approach. Using multiple models allows different leading indicators and model structures to be aggregated into an ensemble forecast, which we have shown to outperform the individual models. This structure allows the ensemble to compensate for limitations in individual models. We validated this model across multiple waves of the COVID-19 pandemic, using proper scoring rules. We show that ensemble models out-score individual models even with a week longer forecast horizon, with the ensemble models having the better interval score and median absolute errors across horizon, wave, and geography.

Throughout the COVID-19 pandemic, hospital forecasting has been an essential part of the public health response worldwide [31] [8] [32]. Models have been developed for a plethora of use cases, from hospital scale workload planning [33] to national level policy making [32]. The range of use cases and challenging landscape of the pandemic have led to a range of methods being developed. Disease transmission models have been widely used for exploring potential scenarios, such as the roadmap out of lockdown in the UK [34]. However, the complex immune landscape and mixture of variants with different transmissibility and immune evasion have led to transmission models becoming increasingly hard to parameterise in real-time, particularly as the sparsity of high-resolution data increases. Therefore, statistical forecasting models have also seen widespread use, such as time series models [35].

Typical time series models for forecasting respiratory healthcare pressures included ARIMA models [36] and deep learning time series approaches [37]. For influenza, as an example, the power in these methods comes from repeated qualitative behaviour across years, where winter surges are seen over similar periods and following similar shapes. However, in the case of COVID-19, regular seasonal behaviour has not yet become established, with waves driven by a mixture of behavioural changes, waning immunity, and viral evolution. Therefore, the main strength of these typical methods is substantially reduced, limiting their forecasting potential. During the 2022/2023 Winter period in England, this reduced performance was also observed for influenza, due to the perturbations to the typical dynamics caused by the COVID-19 non-pharmaceutical interventions [38].

The main limitation in this model is the quality of the leading indicators data. To be used as a leading indicator, syndromic surveillance data must have a stable relationship with hospital admissions. Whilst this has can sometimes been the case [13], variants with a different disease severity profile could cause a step-change in the relationship. Additionally, the quality of syndromic surveillance data relies on individuals in the population mentioning the right terms, which could be biased by public health messaging or the presence of co-circulating infections with similar symptom profiles.

Another limitation comes down the limited number of models included in the ensemble. Future work should focus on combining the individual methods presented here with other forecasting models in an improved ensemble. Additionally, we have only considered a small sample of possible ensembling techniques. More complex techniques, such as Bayesian stacking [39], could lead to stronger ensemble performance.

## Conclusion

In this manuscript we present a set of models used to forecast hospitalisations across the 2022 Omicron waves within UKHSA to guide public health policy. We show ensembling methods can improve epidemic peak performance using purely data driven models and that combining GLMs incorporating leading indicators and hierarchical GAMs with admissions improves predictive performance overall, and over time. We robustly compare predictive performance of the modelling approaches and validate the methods against two Omicron waves. We show that leading indicator models based on leading indicators can help anticipate turning points, but other approaches are can supplement performance in different epidemic phases.

**Conflict of Interest**

The authors have declared that no competing interests exist. The authors were employed by the UKHSA but received no specific funding for this study.

**Data Availability Statement**

UKHSA operates a robust governance process for applying to access protected data that considers:

- the benefits and risks of how the data will be used
- compliance with policy, regulatory and ethical obligations
- data minimisation
- how the confidentiality, integrity, and availability will be maintained
- retention, archival, and disposal requirements
- best practice for protecting data, including the application of ‘privacy by design and by default’, emerging privacy conserving technologies and contractual controls

Access to protected data is always strictly controlled using legally binding data sharing contracts.

UKHSA welcomes data applications from organisations looking to use protected data for public health purposes.

To request an application pack or discuss a request for UKHSA data you would like to submit, contact DataAccess@ukhsa.gov.uk.

**Supplementary Section A**

The exact definition of when an epidemic wave due to a COVID variant is not a single correct date, this also extends to hospital admissions. To discretize the waves and allow comparative analysis we have defined start and end dates for the analysis of each wave in the study. This was done by selecting the lowest points between admissions peaks in England from the UK COVID-19 Dashboard, using the 7-day moving average. While the epidemic peak is of interest, so is the turning point before the growth phase, therefore we include an offset of X days, to capture some time before the turning point. To ensure alignment for weekly forecasts and consistent day-of-weeks, for a start date we select the preceding Sunday, and for the end date the following Sunday. This produces the following start and end dates for each admission wave.

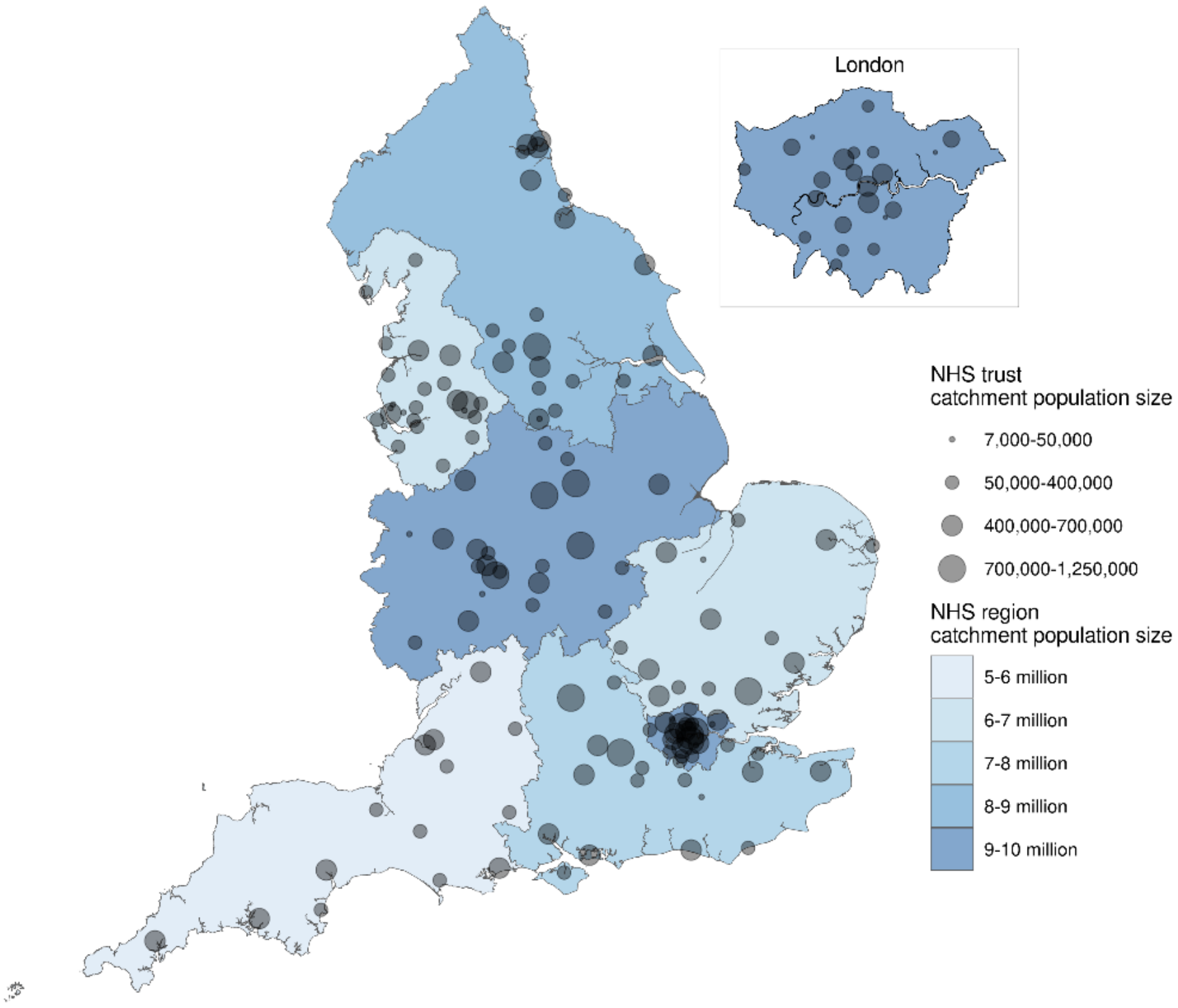

*Supplementary Figure A. Map of NHS regions and trust locations. Population catchment sizes are determined using the local authority weightings derived from historic admitted patients discharge locations. There are a range of sizes of trusts which are often located near each other and therefore covering similar catchment populations.*

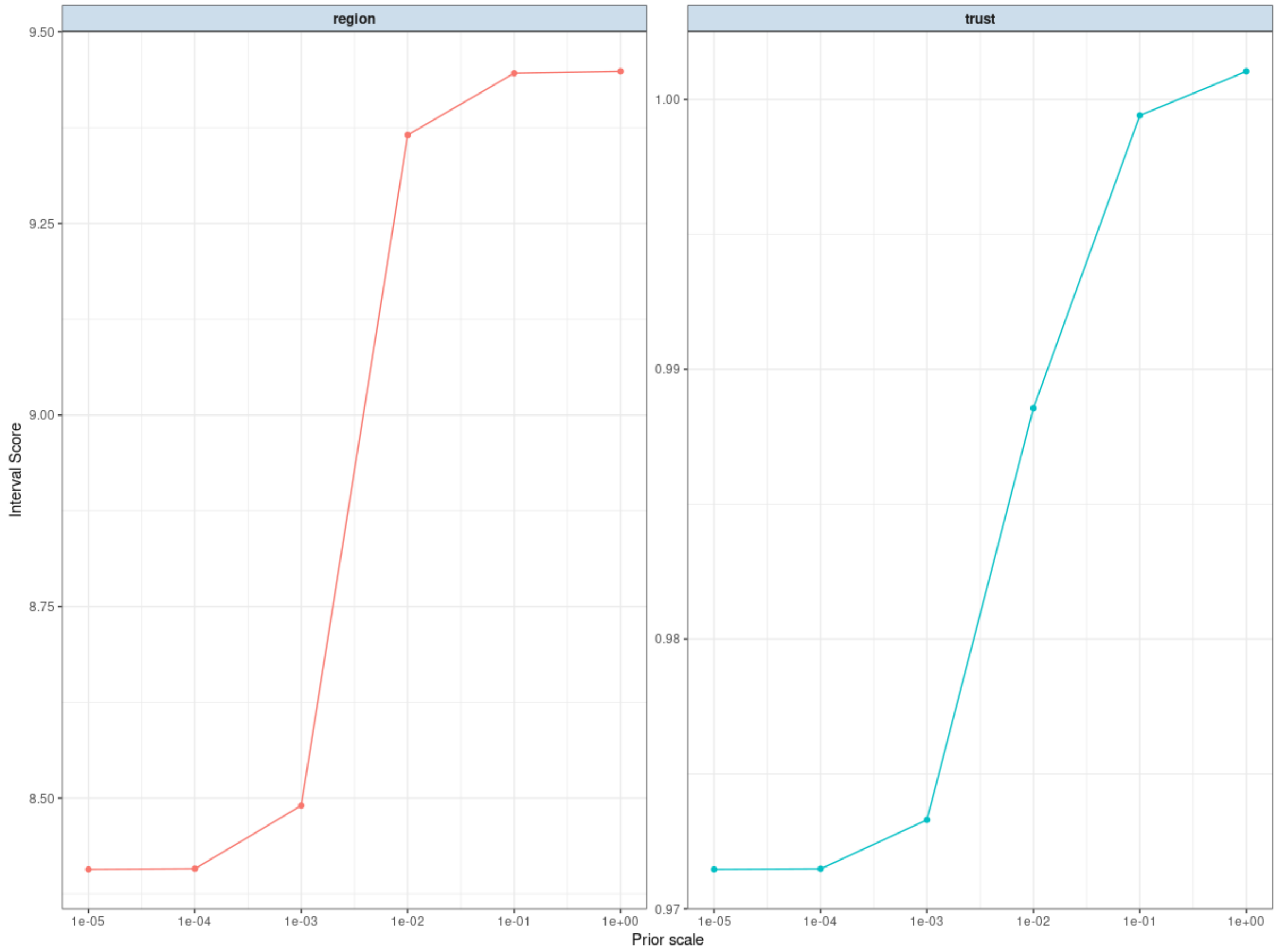

*Supplementary Figure B. Sensitivity with interval score for ensemble regression approach over the BA.4/5 wave across a range of prior scale values with the prior normal(1/n_models, prior scale). A non-informative prior (high value) performs worse than a strong prior (low value).*

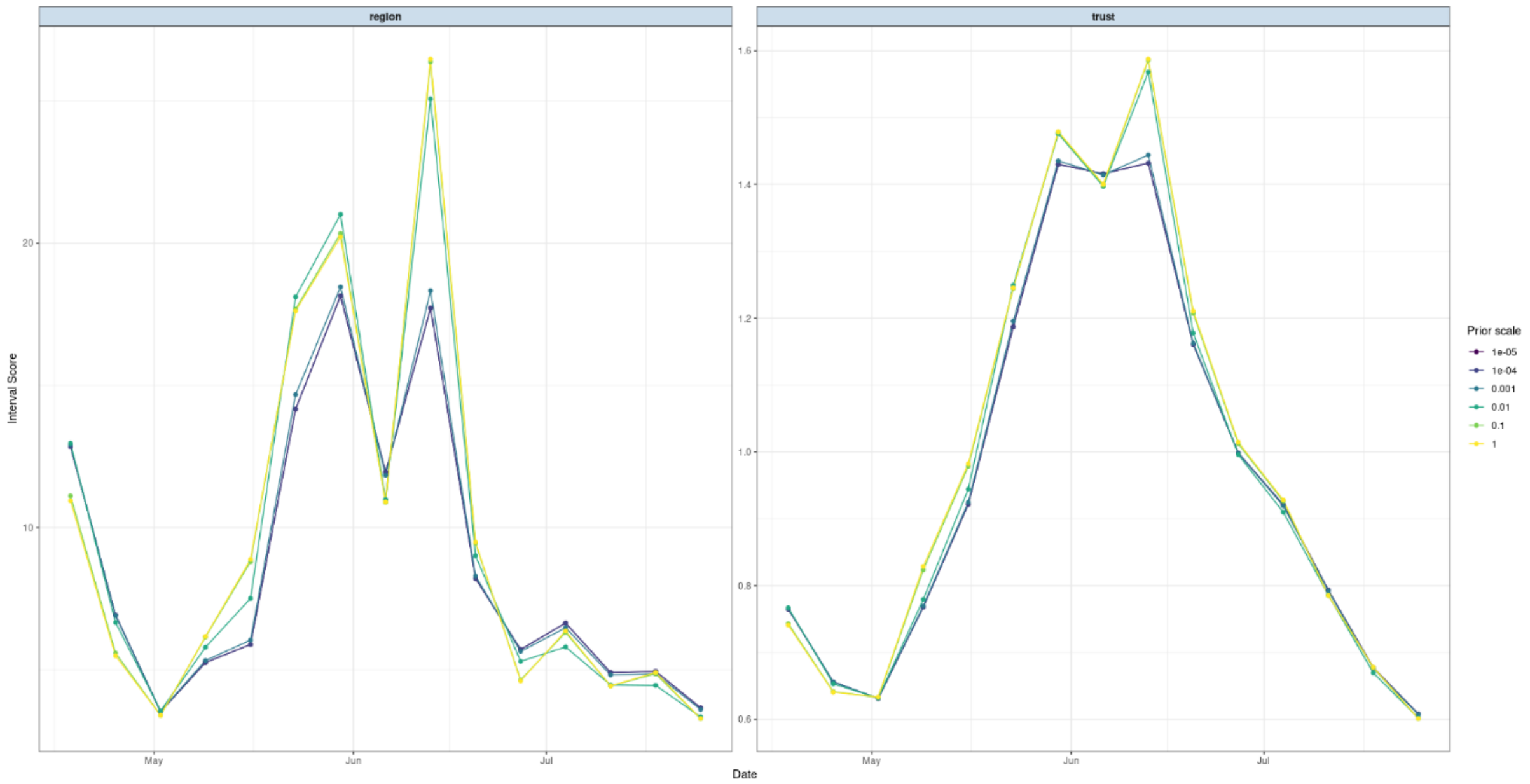

*Supplementary Figure C. Shows the performance of the regression ensemble for the BA.4/5 wave over time. The weaker priors score more poorly at the peak of the epidemic, though similarly elsewhere.*

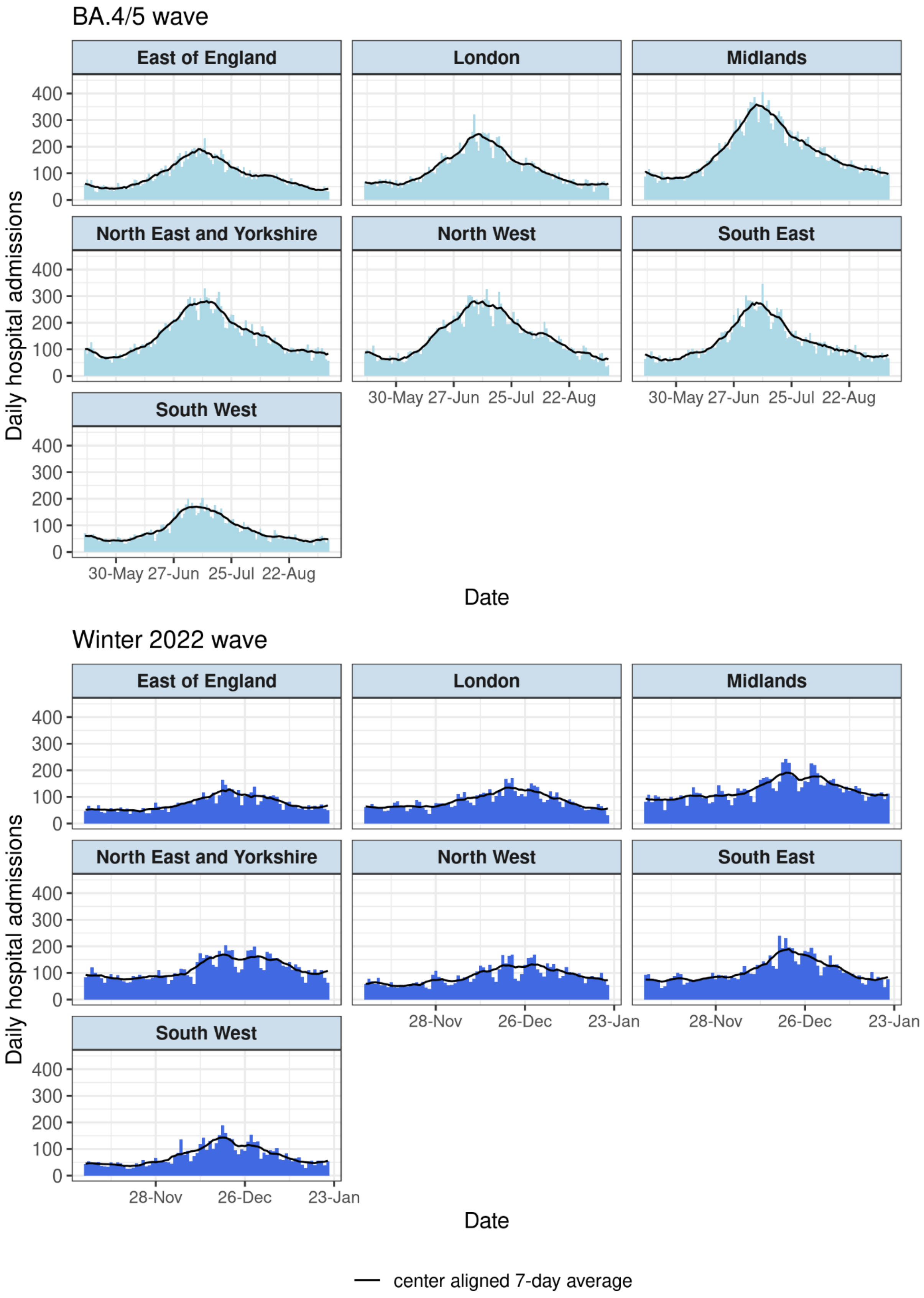

*Supplementary Figure D. Regional stratification of the BA.4/5 and Winter 2022 COVID admission waves.*

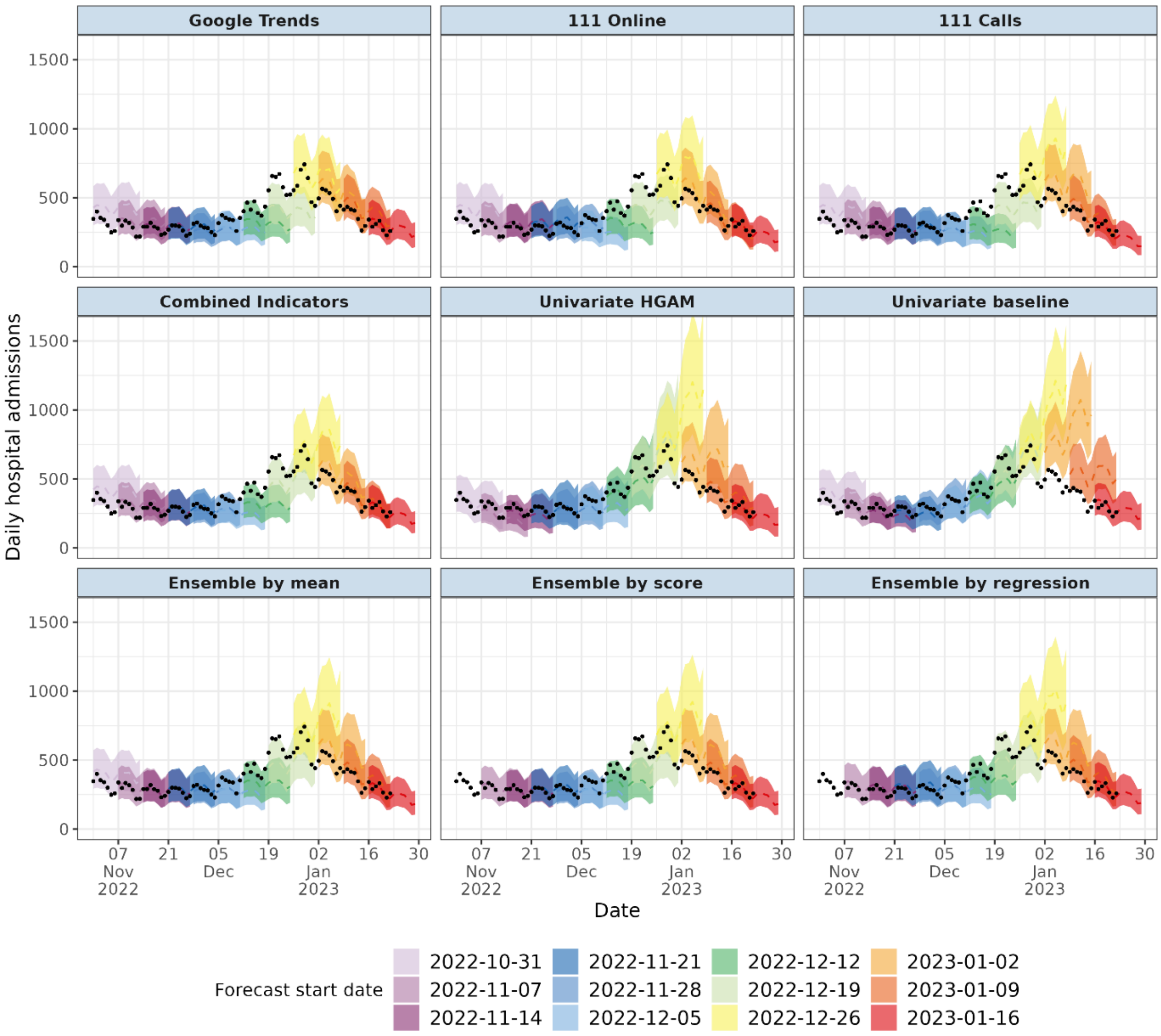

*Supplementary Figure E. The example forecasts of the different model structures for the Winter 2022/23 wave for each week period. The regional forecasts from the GAMs are aggregated to national level to show the epidemic curve and represent forecasts.*

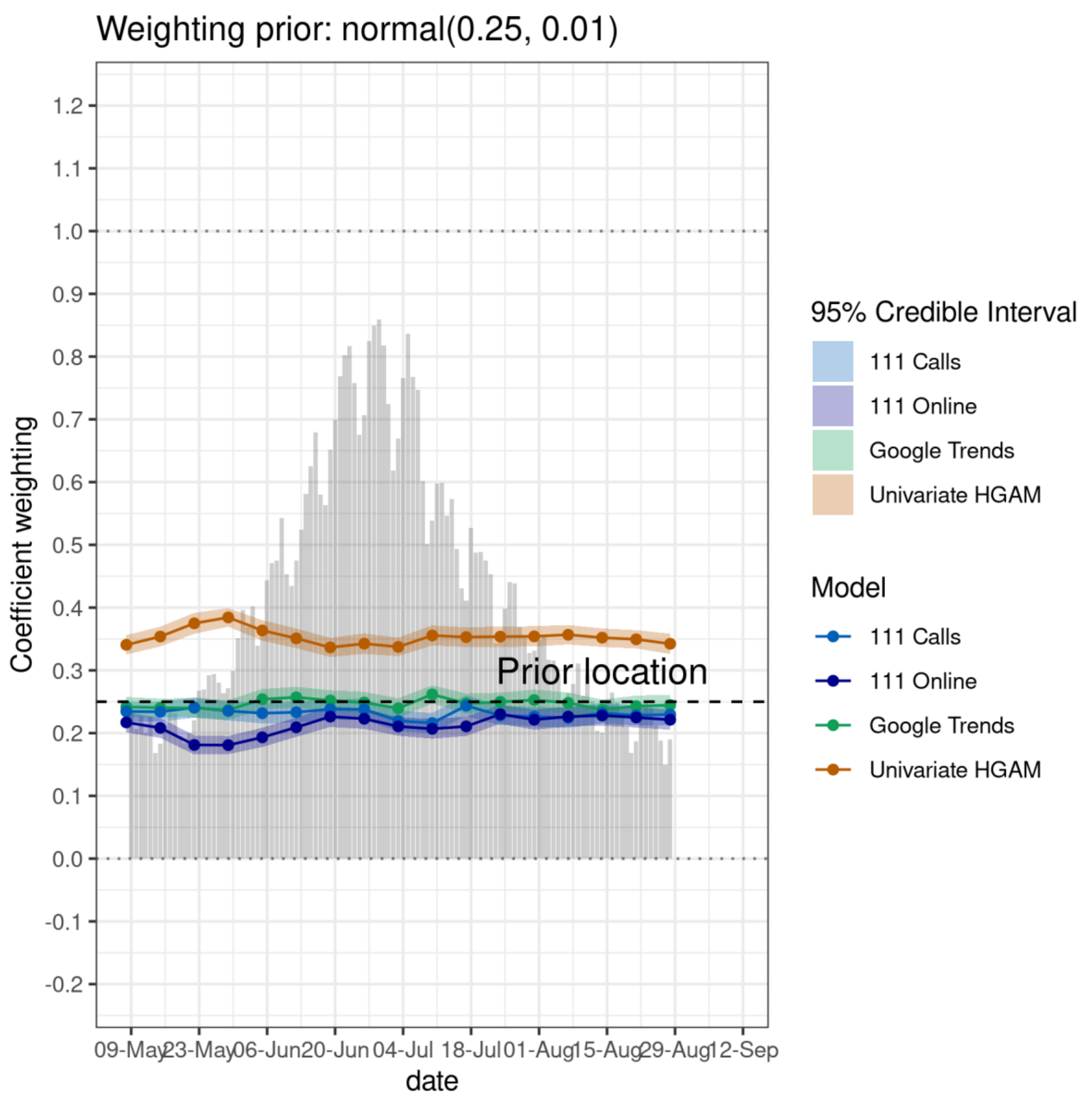

*Supplementary Figure F. The posterior weighting for the ensemble by regression for each model as it changes over time for the BA.4/5 wave. Despite a prior on equal weighting, the posterior weights the Univariate HGAM more highly than other models. This weighting is especially strong during the epidemic growth phase, with higher weighting for the Univariate HGAM and the 111 Online model's weighting is reduced.*

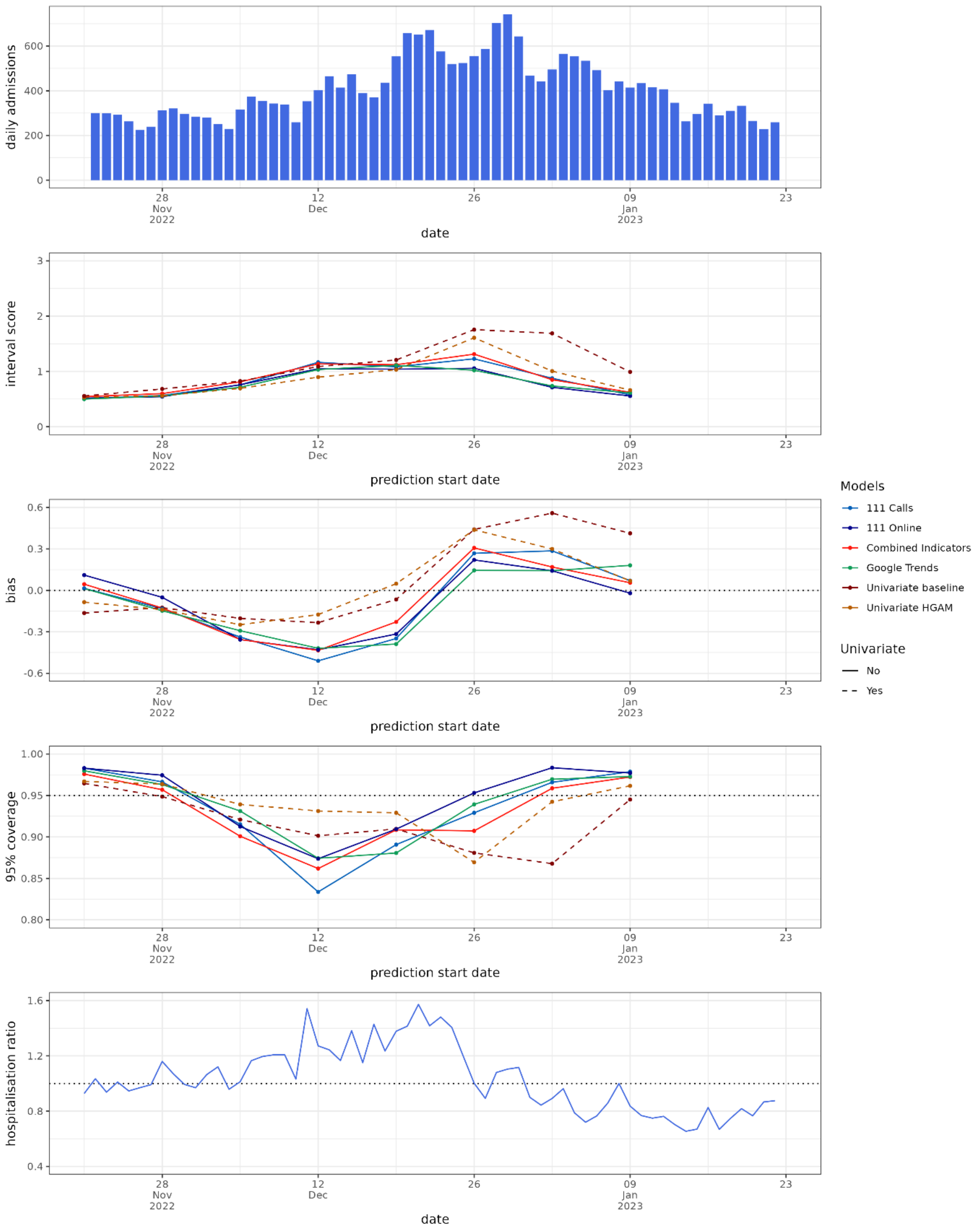

*Supplementary Figure G. Performance of individual models (non-ensembled) over time for the Winter 2022/23 wave. The epidemic curve (top) and hospitalisation ratio (bottom), the admissions divided by the admissions seven days prior, are shown to contextualise scores. The prediction start date represents the first date of prediction, where the predictions will be on the subsequent h=14 days.*

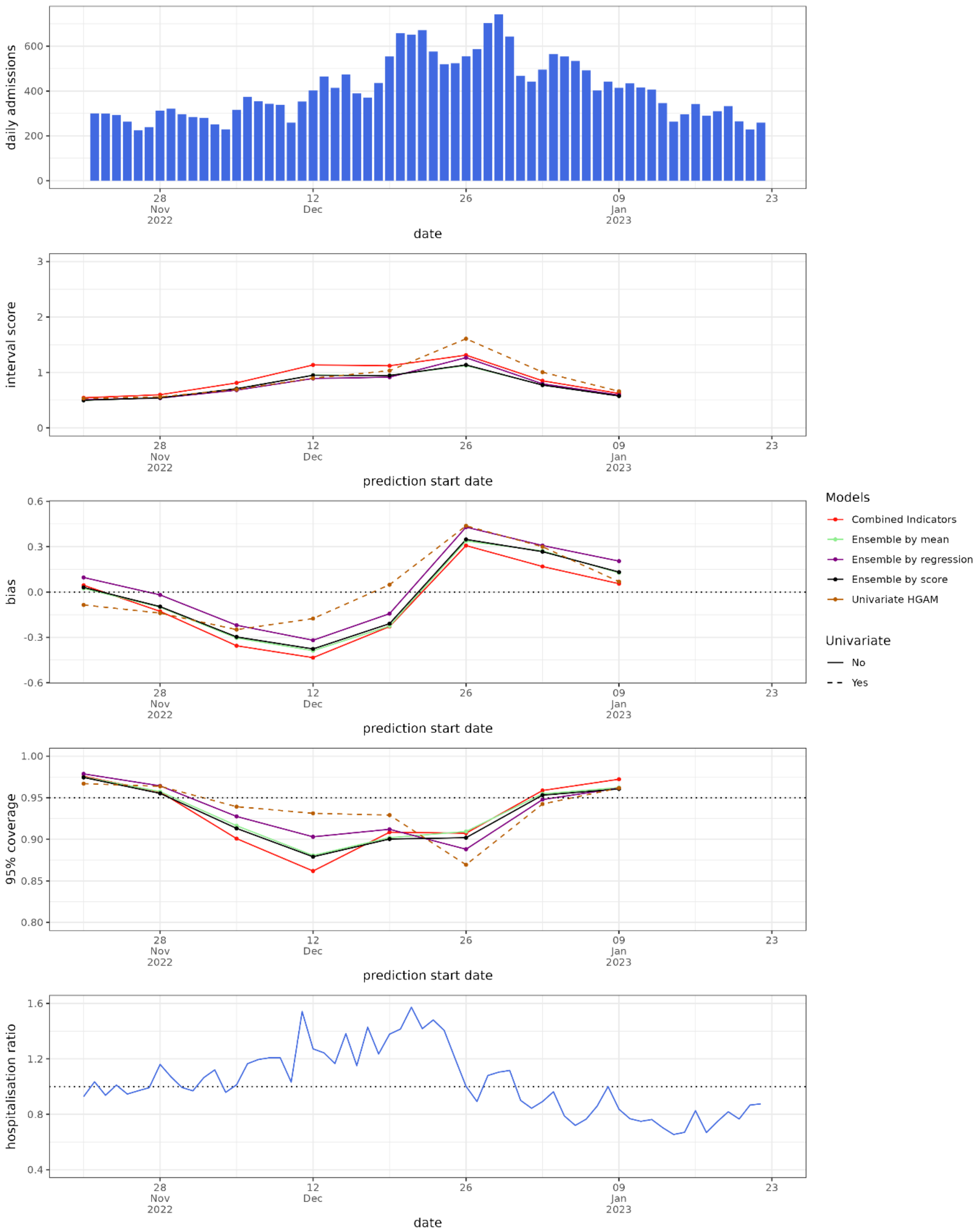

*Supplementary Figure H. Performance of ensemble models over time for the Winter 2022/23 wave, the Univariate HGAM and Combined Indicators model are included to compare performance. The epidemic curve (top) and hospitalisation ratio (bottom), the admissions divided by the admissions seven days prior, are shown to contextualise scores. The prediction start date represents the first date of prediction, where the predictions will be on the subsequent h=14 days.*

| BA.4/5 - Region geography | | | | | | |
|---|---|---|---|---|---|---|
| **model** | **forecast horizon** | **interval score** | **95% coverage** | **median average error** | **underprediction** | **overprediction** |
| Combined Indicators | 7 | 7.13 | 0.803 | 14.9 | 4.21 | 2.11 |
| Univariate baseline | 7 | 9.19 | 0.752 | 18.6 | 2.45 | 5.91 |
| Ensemble by mean | 7 | 5.56 | 0.905 | 12.7 | 2.61 | 2.07 |
| Ensemble by regression | 7 | 5.55 | **0.910*** | **12.6*** | **1.77*** | 2.85 |
| Ensemble by score | 7 | **5.54*** | 0.907 | **12.6*** | 2.56 | 2.08 |
| Google Trends | 7 | 7.42 | 0.786 | 15.4 | 4.61 | **1.99*** |
| 111 Calls | 7 | 7.39 | 0.826 | 16.0 | 3.80 | 2.71 |
| 111 Online | 7 | 6.87 | 0.850 | 14.9 | 3.31 | 2.63 |
| Univariate HGAM | 7 | 6.23 | 0.883 | 14.1 | 2.00 | 3.29 |
| Combined Indicators | 14 | 9.29 | 0.729 | 18.3 | 5.03 | 3.44 |
| Univariate baseline | 14 | 16.5 | 0.658 | 28.6 | 3.51 | 12 |
| Ensemble by mean | 14 | 7.54 | 0.833 | 16.6 | 3.37 | 3.22 |
| Ensemble by regression | 14 | 7.60 | **0.849*** | 16.6 | **2.33*** | 4.26 |
| Ensemble by score | 14 | **7.36*** | 0.840 | **16.3*** | 3.27 | 3.14 |
| Google Trends | 14 | 10.5 | 0.674 | 19.8 | 6.90 | **2.75*** |
| 111 Calls | 14 | 10.9 | 0.7 | 21.1 | 5.45 | 4.52 |
| 111 Online | 14 | 10.1 | 0.75 | 20.1 | 5.13 | 4.09 |
| Univariate HGAM | 14 | 10.1 | 0.815 | 21.1 | 2.89 | 6.05 |
| Combined Indicators | 21 | 14.4 | 0.601 | 25.3 | 8.06 | 5.54 |
| Univariate baseline | 21 | 28.4 | 0.582 | 45.0 | 4.58 | 22.5 |
| Ensemble by mean | 21 | 10.7 | 0.745 | 22.1 | 4.50 | 5.10 |
| Ensemble by regression | 21 | 10.7 | **0.779*** | 22.0 | **3.10*** | 6.47 |
| Ensemble by score | 21 | **10.1*** | 0.758 | **21.2*** | 4.30 | 4.74 |
| Google Trends | 21 | 15.9 | 0.558 | 26.8 | 11.4 | **3.66*** |
| 111 Calls | 21 | 14.6 | 0.628 | 25.9 | 7.32 | 6.36 |
| 111 Online | 21 | 17.7 | 0.555 | 30.3 | 8.80 | 7.94 |
| Univariate HGAM | 21 | 15.9 | 0.783 | 31.3 | 3.98 | 10.3 |

*Supplementary Table A. Scores of each individual and ensemble model across a range of forecast horizons averaged over the BA.4/5 wave, shown for predictions at NHS Commissioning Region level. Best performing models within forecast horizon and metric are denoted with an asterisk (*).*

| Winter 2022/23 - Region geography | | | | | | |
|---|---|---|---|---|---|---|
| **model** | **forecast horizon** | **interval score** | **95% coverage** | **median average error** | **underprediction** | **overprediction** |
| Combined Indicators | 7 | 5.03 | 0.859 | **10.7*** | 2.60 | 1.82 |
| Univariate baseline | 7 | 6.9 | 0.816 | 13.3 | 1.88 | 4.40 |
| Ensemble by mean | 7 | 5.45 | 0.857 | 11.9 | 2.30 | 2.48 |
| Ensemble by regression | 7 | 5.55 | **0.887*** | 12.0 | **1.63*** | 3.22 |
| Ensemble by score | 7 | 5.4 | 0.855 | 11.8 | 2.29 | 2.45 |
| Google Trends | 7 | **4.81*** | 0.866 | 10.2 | 3.03 | **1.18*** |
| 111 Calls | 7 | 8.27 | 0.73 | 16.5 | 3.17 | 4.42 |
| 111 Online | 7 | 5.92 | 0.828 | 12.8 | 2.62 | 2.61 |
| Univariate HGAM | 7 | 5.73 | 0.873 | 12.1 | 1.90 | 3.14 |
| Combined Indicators | 14 | 8.94 | 0.698 | 16.0 | 4.95 | 3.38 |
| Univariate baseline | 14 | 12.5 | 0.731 | 20.8 | **2.26*** | 9.49 |
| Ensemble by mean | 14 | **7.90*** | 0.746 | 15.7 | 3.82 | 3.39 |
| Ensemble by regression | 14 | 8.10 | 0.769 | 16.2 | 2.80 | 4.56 |
| Ensemble by score | 14 | 7.91 | 0.745 | 15.7 | 3.78 | 3.44 |
| Google Trends | 14 | 7.91 | 0.727 | **14.7*** | 5.21 | **2.09*** |
| 111 Calls | 14 | 10.6 | 0.656 | 18.5 | 5.88 | 4.10 |
| 111 Online | 14 | 8.46 | 0.728 | 15.8 | 5.00 | 2.81 |
| Univariate HGAM | 14 | 9.67 | **0.789*** | 18.8 | 2.50 | 6.31 |
| Combined Indicators | 21 | 14.2 | 0.587 | 22.2 | 8.19 | 5.37 |
| Univariate baseline | 21 | 21.6 | 0.672 | 33.9 | **2.56*** | 17.8 |
| Ensemble by mean | 21 | **9.91*** | 0.713 | 18.4 | 4.94 | 4.21 |
| Ensemble by regression | 21 | 10.4 | 0.729 | 19.5 | 3.90 | 5.67 |
| Ensemble by score | 21 | 9.94 | 0.713 | 18.4 | 4.91 | 4.27 |
| Google Trends | 21 | 11.5 | 0.583 | 19.6 | 6.66 | 4.18 |
| 111 Calls | 21 | 11.3 | 0.649 | 18.4 | 8.09 | 2.66 |
| 111 Online | 21 | 10.4 | 0.698 | **16.9*** | 7.42 | **2.39*** |
| Univariate HGAM | 21 | 15.5 | **0.738*** | 28.3 | 3.12 | 11.2 |

*Supplementary Table B. Scores of each individual and ensemble model across a range of forecast horizons averaged over the Winter 2022/23 wave, shown for predictions at NHS Commissioning Region level. Best performing models within forecast horizon and metric are denoted with an asterisk (*).*